# Finding simplicity: unsupervised discovery of features, patterns, and order parameters via shift-invariant variational autoencoders

Maxim Ziatdinov,[1,2,a] Chun Yin (Tommy) Wong,[3] and Sergei V. Kalinin,[1,b]

[1]Center for Nanophase Materials Sciences and [2]Computational Sciences and Engineering Division, Oak Ridge National Laboratory, Oak Ridge, TN 37831

[3]Bredesen Center for Interdisciplinary Research, University of Tennessee, Knoxville

Recent advances in scanning tunneling and transmission electron microscopies (STM and STEM) have allowed routine generation of large volumes of imaging data containing information on the structure and functionality of materials. The experimental data sets contain signatures of long-range phenomena such as physical order parameter fields, polarization and strain gradients in STEM, or standing electronic waves and carrier-mediated exchange interactions in STM, all superimposed onto scanning system distortions and gradual changes of contrast due to drift and/or mis-tilt effects. Correspondingly, while the human eye can readily identify certain patterns in the images such as lattice periodicities, repeating structural elements, or microstructures, their automatic extraction and classification are highly non-trivial and universal pathways to accomplish such analyses are absent. We pose that the most distinctive elements of the patterns observed in STM and (S)TEM images are similarity and (almost-) periodicity, behaviors stemming directly from the parsimony of elementary atomic structures, superimposed on the gradual changes reflective of order parameter distributions. However, the discovery of these elements via global Fourier methods is non-trivial due to variability and lack of ideal discrete translation symmetry. To address this problem, we develop shift-invariant variational autoencoders (shift-VAE) that allow disentangling characteristic repeating features in the images, their variations, and shifts inevitable for random sampling of image space. Shift-VAEs balance the uncertainty in the position of the object of interest with the uncertainty in shape reconstruction. This approach is illustrated for model 1D data, and further extended to synthetic and experimental STM and STEM 2D data. We further introduce an approach for training shift-VAEs that allows finding the latent variables that comport to known physical behavior. In this specific case, the condition is that the latent variable maps should be smooth on the length scale of atomic lattice (as expected for physical order parameter), but other conditions can be imposed. The opportunities and limitations of the shift VAE analysis for pattern discovery are elucidated.

[a] ziatdinovma@ornl.gov
[b] sergei2@ornl.gov

The progress in atomically resolved scanning probe and electron microscopy over the last three decades has opened the pathway to high veracity imaging of atomic structures and electronic phenomena in a broad variety of materials. By now, high resolution (Scanning) Transmission Electron Microscopy has emerged as a primary tool for studies of bulk and 2D materials, capable of visualizing the atomic structure of defects, interfaces, and topological defects associated with chemical and ferroic ordering.[1-6] Similarly, Scanning Tunneling Microscopy and non-contact Atomic Force Microscopy have become primary tools for exploring complex electronic orders in charge density wave and superconducting systems.[7-11]

The emergence of large volumes of the microscopic data at or near atomic resolution necessitates the development of the methods capable of extracting relevant information from large volumes of data. For hyperspectral data sets such as tunneling spectroscopy in STM, force-distance curves in AFM, and electron energy loss spectroscopy (EELS) in STEM, progress was achieved with the introduction of multivariate statistical methods such as principal component analysis[12, 13] and more complex linear dimensionality reduction methods,[14-16] and later simple and variational autoencoders.[17] However, these methods generally apply to 1 (or higher D) spectral data, while correlations in the image plane remain unused and are explored only via human inspection of the corresponding loading/abundance maps.

The advances in machine learning (ML) for image analysis over the last decade have stimulated the interest towards applications of these methods in physical imaging. The supervised ML methods have been broadly incorporated in the analysis of electron and scanning probe microscopy data sets, for example towards atomic and molecular identification.[18, 19] Similar approaches were broadly used for a variety of applications including the search for exotic particles from the Large Hadron Collider data,[20] analysis of satellite images,[21] and cancer detection.[22] However, supervised ML methods explicitly rely on the availability of the labeled data. These typically require human labeling or in special cases using theoretical models.

Over the last year, there has been a progressively growing awareness both in the machine learning and domain communities that applications of supervised ML methods in experimental sciences are associated with a number of unique challenges. Indeed, classical ML applications deal with large data volumes, the proverbial "big data", that sufficiently sample the full distribution of the data. For example, the handwritten digits in MNIST data set or cat images in ImageNet represent a significant subset of all possible writing styles or colorations and poses that cats can adopt, and hence applications to an unknown digit or cat image are in-distribution problems. Comparatively, experimental data sets are sensitive to exact image acquisition parameters and often lead to the out-of-distribution effects, when the supervised algorithm trained for specific conditions will perform poorly for different data acquisition conditions. These issues and some strategies to mitigate these are analyzed in details in Ref [[23]].

These considerations in turn necessitate the development of the unsupervised methods for the analysis of the 2D and hyperspectral images. However, unsupervised methods implicitly rely on inferential biases encoded in the architecture of neural network and loss functions, as well as selection (or engineering) of the feature set.[24] For the neural network architectures, it is generally recognized that classical convolutional neural networks (CNNs) are translationally-equivariant, i.e., during the training they can translate knowledge learned in one location in an image to other locations. At the same time, their response shifts with the position of an object, that is, they do not have a positional *invariance*.[25] As a result, CNNs perform poorly when tasked with learning a representation of arbitrarily shifted objects.[25, 26] The addition of max-pooling layers[27] can partially address this issue if the shifts are (very) small, but not for moderate and large shifts.

An equally important role is played by feature selection. In the absence of any prior data, the images can be represented by a collection of patches centered on the uniform grid that (along with the original position) for a feature set for unsupervised learning. In many cases, imaging data contains physically-defined locations that offer a natural way to define features, e.g., atomic positions in STEM data or center of mass of particles. In these cases, the image patches can be selected centered on these features, and by construction, application of unsupervised ML methods will provide information on local chemical and physical phenomena.[28-33] Here, one of the major limitations is the potential presence of multiple versions of the same object with dissimilar rotations within the image plane, either due to disorder in materials or distortions during the scanning process. The recent introduction of the rotationally-invariant variational autoencoders have allowed mitigating this issue.[34, 35]

However, the availability of physics-defined reference points is not universal to all imaging modes and is directly linked to the signal-formation mechanisms. For example, STEM is preponderantly sensitive to atomic nuclei, and hence contrast maxima provide reliable information on atomic positions. At the same time, STM is sensitive to the electronic density of states. Hence the observed contrast can be expected to have the periodicity of the underlying lattice; however, the observed features can have an arbitrary registry to the lattice depending on local Fermi level position, the proximity of defects, and other factors. Hence, of interest is the separation of the building blocks of such images and remaining factors of variation across the image space that are changing smoothly almost everywhere on the length scale of atomic units and can correspond to physical order parameter fields or microscope distortions.

Here we introduce shift-invariant variational autoencoders (shift-VAE) as a universal method for analysis of such data sets. We pose that the most distinctive elements of the patterns observed in STM and (S)TEM images are similarity and (almost-) periodicity, behaviors stemming directly from the parsimony of elementary atomic structures. This allows for the extension of representation learning – realized here via the shift-VAE – to balance the uncertainty in the position of the object of interest with the uncertainty in shape reconstruction. This approach is illustrated for the model 1D data, and further extended to synthetic and experimental STM and STEM 2D data. The opportunities and limitations of the representation learning analysis for pattern discovery and order parameter field mapping are elucidated.

In representation learning, one assumes that complicated real-world observations originate from a small number of the explanatory factors of variation behind the data.[36] A variational autoencoder (VAE)[37, 38] is a powerful latent variable framework to infer those (latent) factors given the data and *some* prior knowledge. The latter includes a fixed prior distribution over a low-dimensional latent space and, optionally, a small number of labeled data samples.[39] Practically, however, it is usually also necessary to fine tune the VAE's architecture as well as its loss objective for different problems.[24] A VAE consists of the stochastic inference (encoder) and generative (decoder) models parametrized by deep neural networks. The decoder links the latent space to the data space via the likelihood $p(\mathbf{x}|f_\vartheta(\mathbf{z}))$, with $f_\vartheta$ representing a neural network with parameters (weights and biases) $\vartheta$. The encoder maps the data space to the latent space via the posterior distribution $q(\mathbf{z}|f_\varphi(\mathbf{x}))$ where $f_\varphi$ is a different neural network with parameters $\varphi$. This formulation allows turning the VAE approach into an optimization problem where we want to learn the parameters of encoder and decoder neural networks. The loss objective, in this case, is a combination of data reconstruction error and the Kullback-Leibler (KL) divergence term between the encoded $q_\varphi(\mathbf{z}|\mathbf{x})$ and prior $p(\mathbf{z})$ distributions.

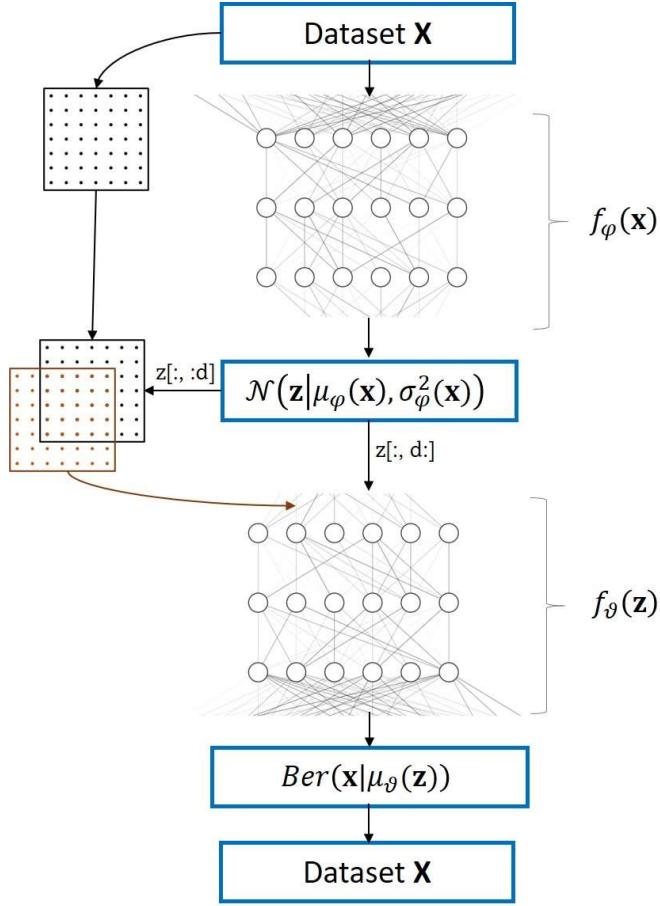

**Figure 1.** Schematic of the shift-VAE. The encoder maps data **x** of dimensionality $d$ ($d$=1, 2) into the mean ($\mu$) and standard deviation ($\sigma$) of the normal distribution from which a latent variable **z** with $n \times (l + d)$ dimensionality is sampled ($l$ is the dimensionality of the conventional VAE latent space). We take the first two columns of the **z** vector ($\Delta\mathbf{r} \coloneqq \mathbf{z}[:, :d]$) and use them to shift our coordinate grid by $k\Delta\mathbf{r}$ where $0 < k \leq 1$ is fixed before the VAE training. The transformed grid is then concatenated with the remaining latent variables ($\mathbf{z} \coloneqq \mathbf{z}[:, d:]$) and passed to a decoder. Finally, we score the observed data against the Bernoulli (*Ber*) likelihood parametrized by the decoder output.

The presence of multiple translated and/or rotated instances of the same objects typically leads to the suboptimal disentanglement of the true *physical* factors of variation in VAE. Indeed, the Galilean principle[40] asserts that mechanical equations of motion are invariant under the $\mathbf{r}' = \mathbf{r} + \Delta\mathbf{r}$ transformation, which is not accounted for by the classical VAE. To incorporate knowledge of basic mechanics into the VAE architecture, we partition our latent space into conventional VAE latent variables and one or two "special" latent variables associated with the variations in object position. The latter is used to shift the 2D grid coordinates of the input data by $k\Delta\mathbf{r}$ where $0 < k \leq 1$ incorporates our "belief" about a degree of disorder in the system. We refer to this modified VAE as shift-VAE and show it schematically in Figure 1. The loss objective is defined as:

$$\mathcal{L} = \mathbb{E}_{\mathbf{x} \sim p_{\text{data}}} \mathcal{L}_{\text{RE}} + \beta(t)\mathcal{L}_{\text{KL}}, \tag{1}$$

where $\mathcal{L}_{\text{RE}}$ is a reconstruction error, $\mathcal{L}_{\text{KL}}$ corresponds to the KL divergence term and $\beta > 0$ is usually introduced to encourage a better disentanglement of the latent representations.[41]

To get insight into the principles and operation of the shift-VAE, we first illustrate this approach for the one-dimensional synthetic data. We define the family of $N = 5000$ sine wave segments with the defined periodicity and varying offsets as $y(x) = \sin((x - \mu)/\sigma) + \sigma_{noise}$ defined on $x \in$ [-10, 10]. The offsets are uniformly distributed on $\mu \in [-3,3]$, whereas periodicities are chosen to have (small) discrete number of values. This choice is made both to explore the relationship between the factors of variability in data and VAE performance, and to approximate an experimental scenario when only few possible wavelengths are present in a single image. White noise of amplitude $\sigma_0$, or amplitude uniformly distributed on $\sigma_{noise} \in$ [0, 0.1] interval, can be added. Examples of the data set are available through the accompanying Google Colab notebook.

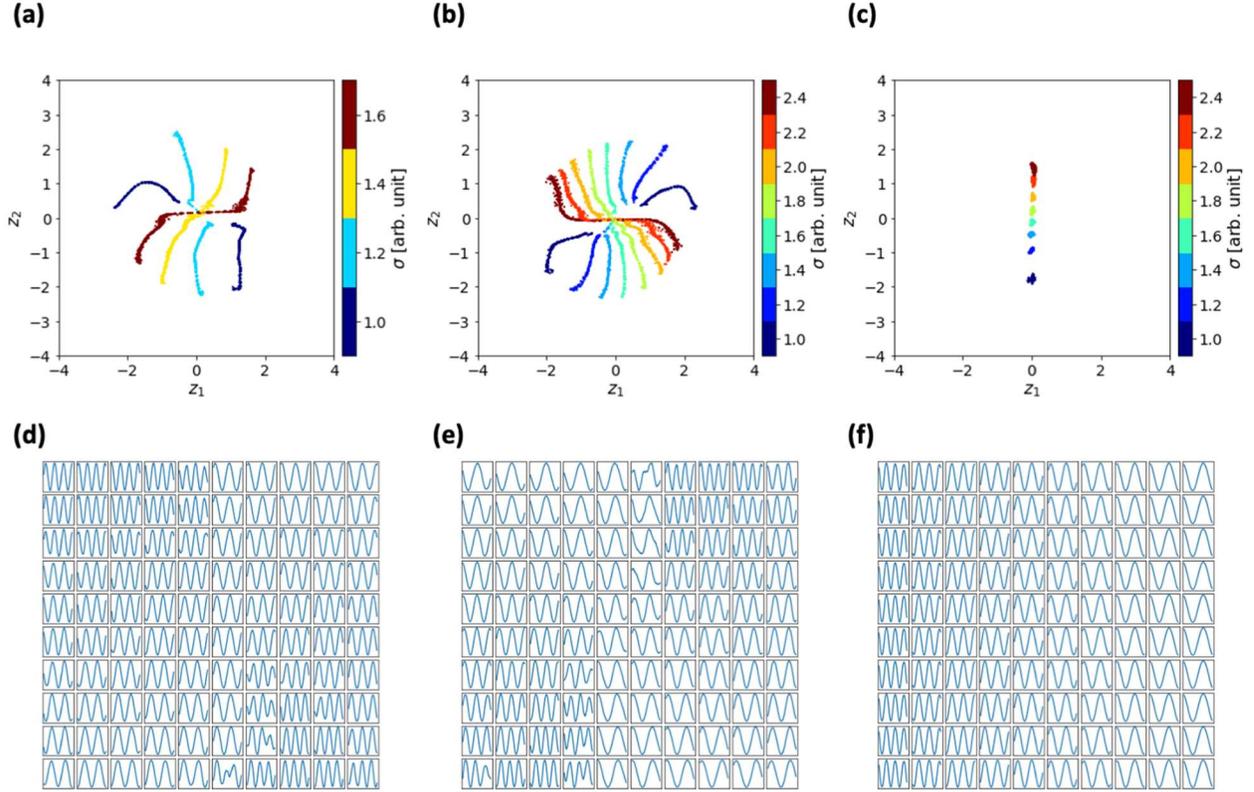

**Figure 2.** Latent space distributions formed by the data encoded with a regular VAE for (a) $\sigma \in [1, 1.6, 0.2]$ and (b) $\sigma \in [1, 2.4, 0.2]$. (c) Latent space distribution formed by the data encoded with the shift-VAE for $\sigma \in [1, 2.4, 0.2]$. (d-f) Corresponding learned latent space representations projected to the data space. Here, the distribution of periodicities is defined as $\sigma \in [\sigma_{min}, \sigma_{max}, \sigma_{step}]$.

The analysis of the data using a regular VAE is shown in Figure 2 (a,b) for several distributions of periodicity. In this case, the encoded data shows a set of well-formed 1D manifolds in the latent space, with the phase changing uniformly across the manifold. Note that the manifolds corresponding to different periods are not necessarily continuous and can be split into two or more. These behaviors can be understood as the result of the countervailing tendencies of VAEs to represent the data in the compact form and at the same time from the fact that manifolds for different periodicities cannot cross (since cross-point will be unphysical). These trends are even more pronounced for the larger number of periodicities, as shown in Figure 2 (b). It is also important to note that these patterns are closely related to the sampling of parameter space, i.e., a large number of possible offsets and a small number of periodicities (but this, in turn, is chosen to comport to plausible experimental scenarios).

Note that despite the fact that for a simple VAE the latent variables do not allow direct interpretation, they nonetheless allow exploring the characteristic features of data. Here, the postulated premise is that real materials will have only a small set of atomic spacings as imposed by preferred bond lengths. In the presence of noise or a more complex structure, these manifolds will broaden. The corresponding projections of the latent space representations to data space are shown in Figure 2 (d,e), and visualize data decoded from the latent points distributed over the uniform spatial grid. The reconstructed objects clearly illustrate sine waves, and these behaviors are further analyzed later.

In comparison, shown in Figure 2 (c) is the latent space for shift-VAE. Here, one of the latent variables is an offset, and the remaining two are the classical continuous latent variables. Note that in this case the manifolds corresponding to different periodicities are clearly separated. Furthermore, the latent space is clearly quasi-collapsed, i.e., data points are distributed only along one latent direction. While typically perceived to be a problem for VAEs, we pose that this behavior in fact allows inferring the physics of the system, namely the number of the physical factors of variability. In this case, the synthetic data set has two factors of variability – phase and periodicity, with the phase separated as a shift. Correspondingly, in this case, periodicity becomes associated with one of the latent variables, and the second latent variable collapses, in agreement with this conjecture.

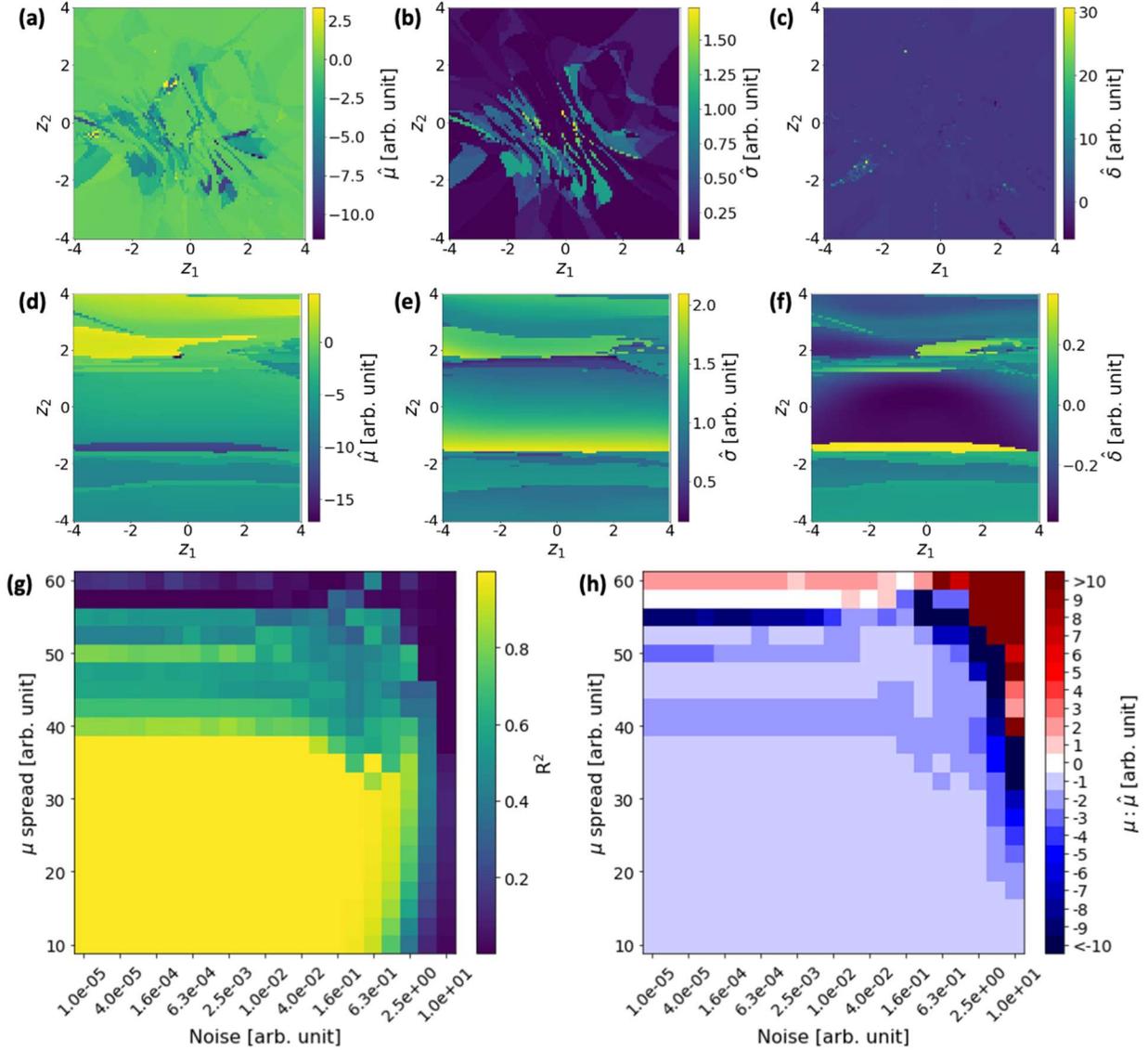

**Figure 3.** The sinusoidal fit parameters in the VAE latent space. Shown are maps for (a) fitted shift, $\hat{\mu}$, (b) fitted period, $\hat{\sigma}$, and (c) fitted amplitude, $\hat{\delta}$, respectively. Fitted parameter maps in shift-VAE latent space for (d) fitted shift, $\hat{\mu}$, (e) fitted period, $\hat{\sigma}$, and (f) fitted amplitude, $\hat{\delta}$. Compared to the VAE latent space, the shift-VAE latent space shows several distinct linear regions corresponding to the transition of the fit parameters. (g) and (h) show the error in the fit in terms of $R^2$ and the ratio $\mu:\hat{\mu}$, delineating the region where shift-VAE latent shift coincides with ground truth value.

We further explored the structure of the latent space for the VAE and shift-VAE, namely the relationship between the latent parameters and physical descriptors. Specifically, we decoded the signal generated with parameters $\mu \in [-3,3]$, $\sigma \in [1, 2.2, 0.2]$ from each point on the latent space grid and fit it to the sine wave. The fitting function is defined as $y(x) = \hat{\delta}\sin((x-\hat{\mu})/\hat{\sigma}) +$ $\hat{\sigma}_{noise}$, where $\hat{\delta}$ is the amplitude and the hat indicates fitted parameters as opposed to parameters used to generate the data.

The fit parameters, $\hat{\mu}$, $\hat{\sigma}$, and $\hat{\delta}$, at each pair of latent parameters are shown in Fig. 3 (a)–(c) for VAE and (d)–(f) for shift-VAE. Note that for simple VAE the parameters in latent space are distributed following complex patterns, as can be expected from original data distributions shown in Fig. 2 (a,b). At the same time, Fig. 3 (d)–(f) illustrate the efficacy of the shift-VAE on separating sine curves with different fit parameters. The efficacy, however, drops sharply with increasing variances in the fit parameters. In such a manner, we establish the relationship between pairs of encoding latent variables $(z_1, z_2)$ and the parameters $\hat{\mu}$, $\hat{\sigma}$, and $\hat{\delta}$. We can also plot the fit error, defining the discrepancy between the data and the ground truth functional form. Note that for shift-VAE the expected error for offset variable should be zero by construct, and hence deviation from zero provides a measure of the disentanglement of shift variable, as will be explored below.

Finally, we note that the convergence of the shift-VAE strongly depends on the distribution of the initial data. For small and moderate phase offsets, the recovered shift value is equal to the ground truth offset. For larger offsets, the behavior starts to deviate from the ground truth value. Furthermore, in some cases the training process leads to the emergence of the metastable minima, i.e., the manifold for the single periodicity will be split in two as shown in Fig. 3 (d)–(f).

To explore these behaviors systematically, shown in Figure 3 (g) and (h) are the $R^2$ and the ratio $\mu:\hat{\mu}$ for pairs of $\mu$ spread and $\hat{\sigma}_{noise}$ values. Here, $\mu$, $\sigma$, and $\delta$ are varied independently from $\mu \in [-6,6]$, $\sigma \in [1.8,12]$, and $\delta \in [1,12]$. The error in the sine fit is then evaluated by first performing linear regression on the latent variable against the fitted parameter and evaluating both the $R^2$ coefficient for the quality of the fit, and the ratio between the latent variable and the fitted parameter. Here, the ratio greater than 1 means that the latent variable overpredicts the fit parameter of the sine wave, and a value less than one means that the latent variable underpredicts the fit parameter. Worth noting is the sharp decrease in the quality of the fit for $\mu > 5.6$ and $\hat{\sigma}_{noise} > 1$ in Fig. 3 (g), meaning that outside the yellow region the shift-VAE disentanglement of the shift variable fails. Similarly, Fig. 3 (h) reveals that there is a gradual underprediction of the latent variable before a sharp overprediction of the latent variable, resulting in a high instability at the boundary. We further note that this behavior can be controlled through the choice of the priors on the variational distributions and the loss function, and an approach for this will be illustrated below.

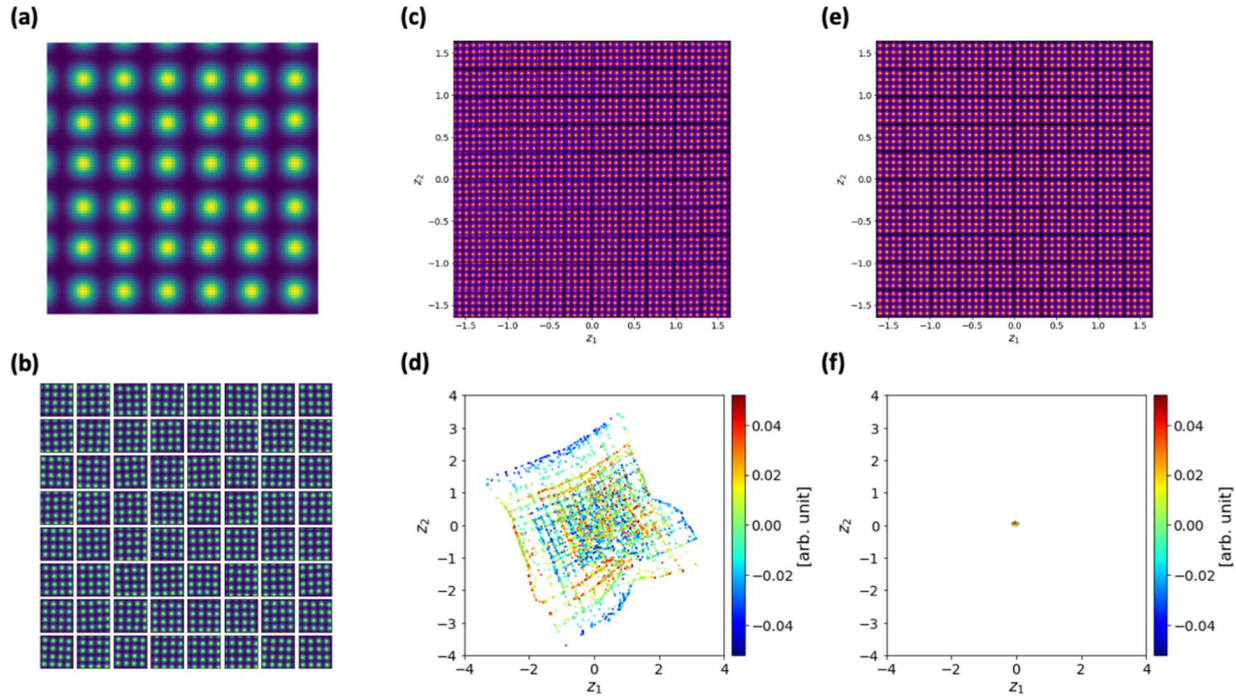

**Figure 4**. (a) Synthetic atomic-resolution image. The positions of the atoms are corresponding to ideal periodic lattice with the added Gaussian positional noise. (b) The representative data set of patches with the sampling points chosen with uniformly distributed random offsets relative to lattice position with an angle of rotation between -2° and 2°, and a displacement of atomic position ∈ [0,0.1]. Hence, sampling points contain only partial information on atomic positions. (c) The latent space and (d) latent distribution for VAE. Note the systematic evolution of offsets across the latent space and complex latent distribution. (e) Latent space and (f) latent distribution of shift-VAE. Note that latent space is fully collapsed, since all factors of variability (x- and y- offsets) are accommodated by the special offset latent variables of the shift-VAE.

We further extended the shift-VAE approach to a 2D case. In this case, we introduce two special latent variables designated to encode the information about the offsets, giving rise to 2+$l$ latent space. For convenience of representation, we choose $l = 2$. As a synthetic data set, we use a simple square lattice with ideal translational symmetry and a 2D Gaussian centered at each lattice site as shown in Figure 4 (a). To create image patches, the sampling points are chosen with uniformly distributed random offsets with respect to ideal lattice position, as shown in Figure 4 (b). Here, if the distribution is very narrow, this implies that all the patches are centered at atomic units. If the distribution is broad (or sampling points are chosen randomly), the patches do not contain information on atomic positions.

The simple VAE analysis of thus created data set with an angle of rotation between -2° and 2°, and a displacement of atomic positions ∈ [0,0.1] is illustrated in Figure 4 (c,d). Here, the projected latent manifold of the VAE is populated by the image patches with different offsets. The corresponding encoding of the data into the latent space shown in Figure 3 (d) has a very complex structure formed by the multiple linear manifolds. The emergence of this structure is attributed to the digitization effects, i.e., a relatively small number of the pixels in the image patches. However, we note that the conditions here are chosen to be close to experimental, and in the subsequent discussion we will see that the similar behavior will manifest in the experimental data. The color represents the displacement of the atomic position.

In comparison, the latent space of the shift-VAE projected to the data space appears to be populated by identical elements (Fig. 4e). Correspondingly, the associated latent distribution is highly localized and both continuous latent variables are collapsed (Fig. 4f). Similar to the example of 1D data set, we attribute this behavior to the fact that offsets were the only physical factors of variability in the original data set. Here, the offsets separated as two special latent variables are the only two factors of variation and it is unsurprising that the remaining variability is absent and latent space is doubly degenerate.

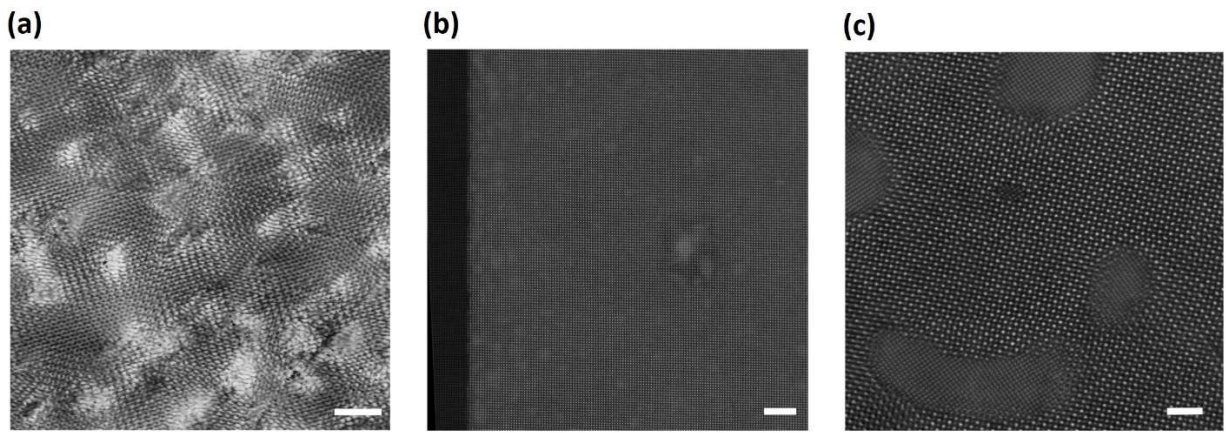

**Figure 5.** The experimental images used in the study. (a) STM of graphene oxide.[42] (b) STEM of Sm-doped BiFeO$_3$[43], (c) STEM of NiO-LSMO.[23] The scale bars are (a) 1 nm, (b) 6 nm, (c) 2 nm.

With these analyses in hand, we extended this approach to the experimental data sets. As models, here we use the STM images of graphene (Fig. 5a), and STEM images of BiFeO$_3$ (Fig. 5b) and NiO-LSMO heterostructure (Fig. 5c). These datasets have been used in prior publications [23, 43, 44], which give the details on the nature of the system and imaging conditions. Here, they are used as representative examples of systems with complex spatially-varying patterns with atomic resolution. The selection of these three systems is predicated by the nature of anticipated behaviors, namely perturbation of electronic structure of graphene on a large number of individual defects, presence of two phases, ferroic domains, and chemical variability in BiFeO$_3$, and two-phase coexistence in the NiO-LSMO composite.

Similar to the synthetic data, the images are transformed into collection of patches sampled over the square grid. The window size and the grid sampling can be varied to match the specific physics of the image. Here we have found that optimal performance can be obtained for patch size incorporating 2-3 unit cells, with the grid step being ½ of the patch size. Note that the grid spacing in this case is chosen randomly and does not contain any information on atomic positions. The provided Google Colab notebook allows readers to experiment with these parameters or analyze own data. However, it is important to note that the positions of the atomic units (as used previously in Ref [45, 46]) in this case are not available and the choice of the rectangular grid represents lack of prior knowledge or physical inferential biases. We also note that for graphene data we found that each patch should be normalized; otherwise, the relative intensity changes become the primary factor of variation in data.

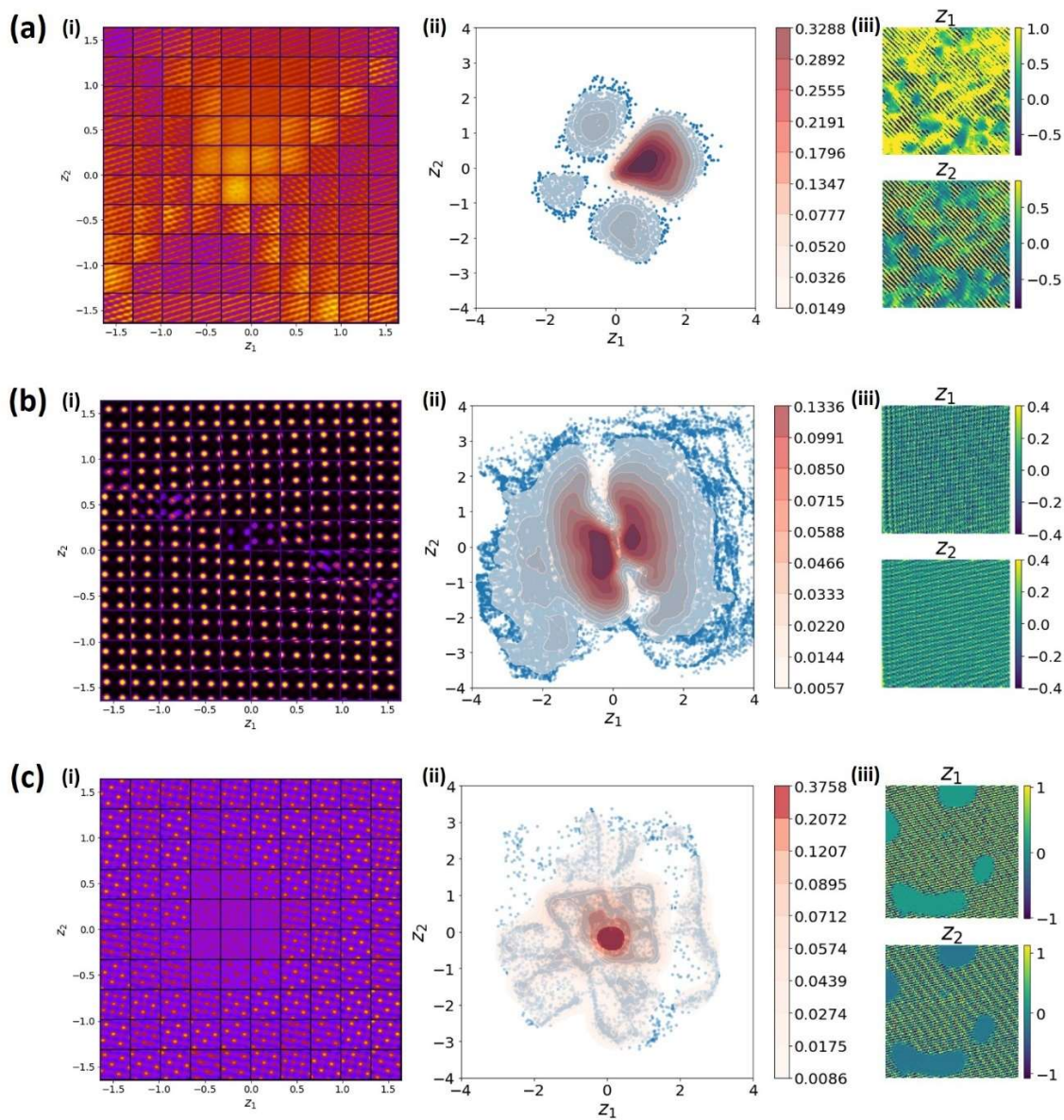

**Figure 6.** The regular VAE analysis of the experimental data:(i) the learned latent manifolds projected to the data space, (ii) the encoded latent distributions, (iii) and (iv) the latent variables, $z_1$, $z_2$, in the image spaces for (a) graphene, (b) BiFeO$_3$, and (c) NiO-LSMO.

The VAE analysis for chosen systems is illustrated in Figure 6. In all cases, we show the (i) learned latent manifolds decoded to the data space, (ii) encoded latent distributions, and (iii,iv) spatial maps of the encoded latent variables. Note that compared to the idealized model system in Figure 4, these images represent both partially known physics of these materials, and also contain the non-idealities of the real imaging system.

For graphene, the learned latent space representation (Fig. 6a-i) contains elements resembling the ideal hexagonal graphene lattice, and elements resembling waves in one of the chosen lattice directions. The distribution of these in the latent space is fairly complex, as is typical for the system having rotational degeneracies (which in this case are not captures by the latent variable). The corresponding encoded distribution (Fig. 6a-ii) is relatively simple and contains a small number of crisscrossing features. The spatial latent variable maps clearly separate the defect and non-defect regions and are dominated by the high-periodicity features formed by the beating between the sampling square lattice and intrinsic material lattice. For $BiFeO_3$, the learned latent space is populated by the distorted versions of the perovskite unit cell (Fig. 6b-i), whereas for NiO-LSMO system careful examination illustrate presence of A- and B-site centered blocks and rocksalt blocks (Fig. 6c-i). The corresponding encoded latent distributions (Fig. 6b-ii and 6c-ii) are very complex, resembling that in Fig. 4 (d). The corresponding spatial maps are again dominated by the beating effects.

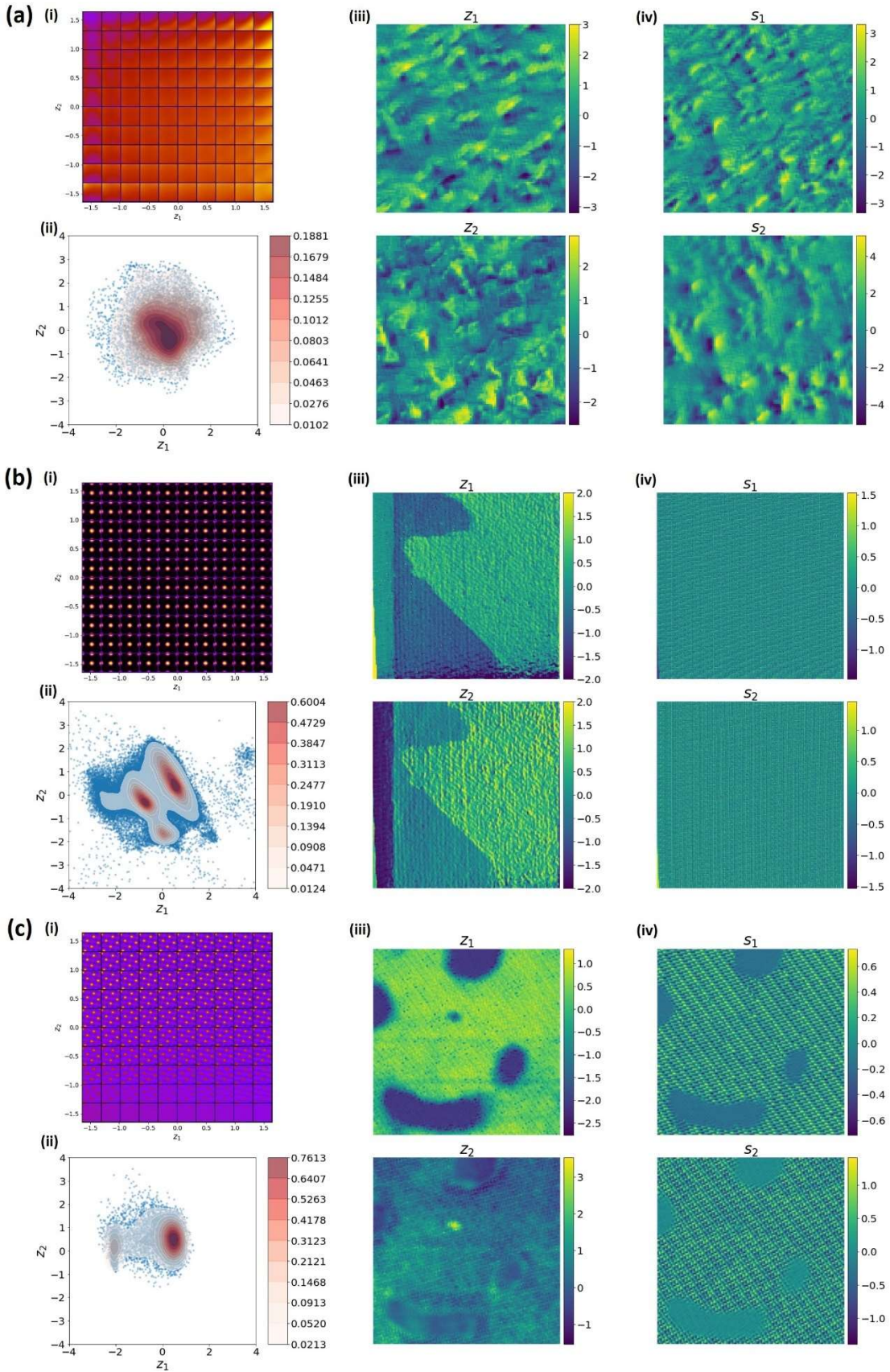

**Figure 7.** The regular VAE analysis of the experimental data: (i) the learned latent manifold projected to the data space, (ii) encoded latent distributions, (iii) the conventional ($z_1$, $z_2$), and (iv) offset ($s_1$, $s_2$) latent variables in the image spaces for (a) graphene, (b) BiFeO$_3$, and (c) NiO-LSMO.

In comparison, the corresponding shift-VAE analyses are shown in Figure 7. Here, shown are (i) the learned latent manifolds projected to the data space and (ii) the encoded latent distributions as well as the spatial maps corresponding to (iii) two conventional latent variables and (iv) two offset latent variables. Note that in this case the learned latent manifolds are formed by identical or weakly-changing blocks, whereas the encoded latent distributions are particularly simple and formed by the small number of modes. By inspection of the spatial maps, these can be identified with the substrate and ferroic variants in the BiFeO$_3$ system (Fig. 7b) and the phases in the NiO-LSMO system (Fig. 7c).

In the corresponding spatial maps, the latent shifts show the beating pattern as can be expected, whereas the latent variables show smooth maps. Here, we note that this is the required behaviors, since typically order parameters in physical systems are postulated to change smoothly on the length scale of the unit cell.

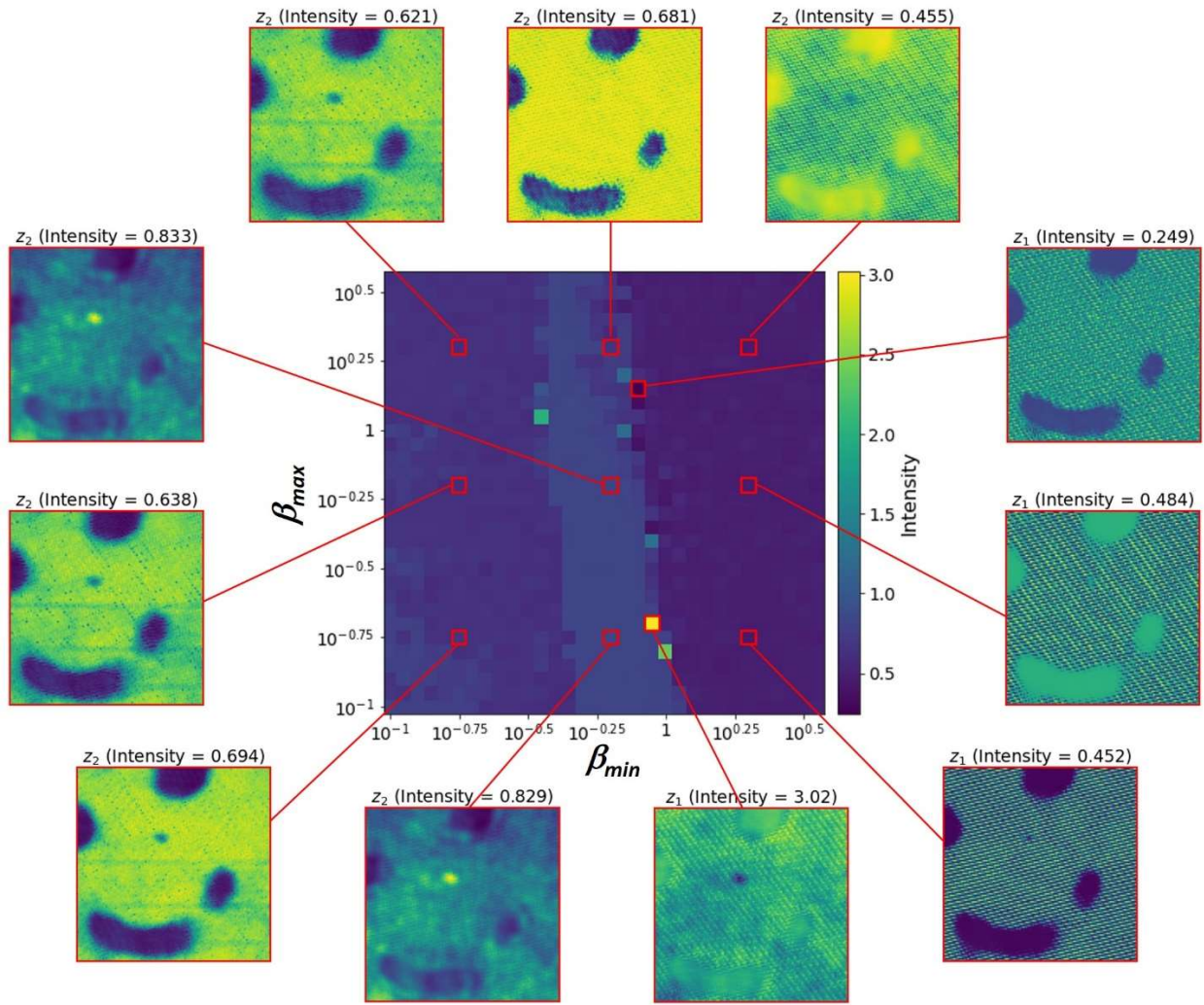

**Figure 8.** Grid search for the optimal KL scale factor ($\beta$) and resultant values of the latent variables in the image space for the selected grid points.

Finally, we note that for a broad offset distribution, the shift-VAE can converge to multiple minima, and under certain conditions the high-frequency periodic features can "leak" from the offset latent variables to conventional latent variables. We found empirically that this process can be controlled by the relative weights of the reconstruction and KL terms expressed via a scale factor $\beta$ in Eq. (1). We further systematized this analysis by exploring the effect of this parameter as a function of time to control the shift-VAE training.

We note that to find an optimal KL scale factor for obtaining physically relevant latent representations, one must introduce a "physicality" parameter defining how the result should "look like". Here, we take advantage of the fact that an order parameter field has to be smooth on the length scale of atomic units. This means that latent variables change smoothly over the image, whereas latent offset images show periodicity corresponding to the Moire pattern between the sampling lattice and intrinsic atomic lattice of materials.

To implement this, a grid search is carried out for KL scale factors ranging from 0.1 to 10 for initial and final KL value, respectively. The Fourier transformation is computed for the resultant latent variables, $z_1$ and $z_2$, in image spaces. The ratio of central peak intensity to the total image intensity yields the measure of smoothness of the data since the periodic features yield the peaks at finite values of the reciprocal lattice vectors. The greater of the intensities computed from the two latent variables is used to generate the heat map as a measure of the performance of the shift-VAE.

The resultant dependences are shown in Figure 8. Here, the final value of the KL scale factor affects the resultant latent maps only weakly, whereas initial value has a strong effect on convergence of the shift-VAE. For small initial values of the KL scale factor, the resultant maps are smooth, i.e., comport to the behavior expected to physical order parameter. For larger KL values one of the maps starts to exhibit the periodic features due to the beating between sampling and the atomic lattices, and for even large parameter values both maps contain these high frequency features. It is important to note that within these individual regions the maps are generally similar, i.e., once the right region of the hyperparameter space is identified the answers (for the "smoothness" parameter) does not depend on exact KL value.

This behavior can be rationalized from an information theory point of view. The larger penalties on the KL term result in a significantly restricted capacity of the latent bottleneck. In this type of scenario, the optimal behavior of (shift-)VAE is to encode only the positional information since it can provide the most gains for data log-likelihood (reconstruction term in Eq. 1) compared to other factors of variation. This forces the shift-VAE to start encoding the information about shifts into a conventional latent variable instead of the special offset latent variable leading to the aforementioned leakage. By starting with a smaller KL scale factor and gradually increasing it, we encourage our model to encode other (than shift) factors of variation into the conventional latent variables and force the learned latent factors to be more independent, while preserving a high-quality data reconstruction (not shown).

To summarize, we introduced a shift-VAE approach for the discovery of the repeated features and patterns in images. This approach can be used for physics discovery in atomically resolved scanning transmission electron microscopy and scanning tunneling microscopy data and is made possible by the parsimony of elementary atomic structures. Importantly, it allows separating the individual building blocks present in a small number of discrete classes from the relevant order parameter fields. For the latter, we assume that they (almost everywhere) change slowly on the length scale of atomic lattice.

We further demonstrated that the shift-VAE training can be guided though the optimization of the KL scale factors, allowing to incorporate the inferential biases into the system. We note that in the online settings (i.e., during the actual STM/STEM experiments) the grid search strategy used here may not be the optimal one. In the future, the utilization of the Bayesian optimization techniques[47, 48] may lead to a better sample efficiency and faster convergence to a physically-meaningful latent space.

**Acknowledgements**: This effort (ML and STEM) is based upon work supported by the U.S. Department of Energy (DOE), Office of Science, Basic Energy Sciences (BES), Materials Sciences and Engineering Division (S.V.K.) and was performed and partially supported (M.Z.) at Oak Ridge National Laboratory's Center for Nanophase Materials Sciences (CNMS), a U.S. Department of Energy, Office of Science User Facility.